\documentclass[runningheads]{llncs}

\usepackage{psfig}
\usepackage{alltt}
\newcommand{\objectname}[1]{\mbox{@#1}}
\newcommand{\predicate}[3]{\item[#1({\it #3})]\mbox{}\hfill\\}
\newcommand{\predref}[2]{{\bf #1/#2}}
\newcommand{\postscriptfig}[3]{%
        \begin{figure}%
        \centerline{\psfig{file=#2.eps,#1}}
        \caption{#3}
        \label{fig:#2}
        \end{figure}}
\newcommand{\secref}[1]{section~\ref{sec:#1}}
\newcommand{\tabref}[1]{table~\ref{tab:#1}}
\newcommand{\figref}[1]{figure~\ref{fig:#1}}
\newcommand{\Figref}[1]{Figure~\ref{fig:#1}}
\newcommand{\var}[1]{{\it #1}}
\newcommand{\class}[1]{{\bf #1}}
\newcommand{\type}[1]{{\it #1}}
\newcommand{\ctype}[1]{{\it #1}}

\begin{document}
\setcounter{page}{97}
\title{An Architecture for Making Object-Oriented Systems Available from Prolog}
\titlerunning{Making Object-Oriented Systems Available from Prolog}
\author{Jan Wielemaker \and Anjo Anjewierden}
\authorrunning{J. Wielemaker, A. Anjewierden}
\institute{Social Science Informatics (SWI), \\
           University of Amsterdam, \\
           Roetersstraat 15, 1018~WB Amsterdam, The Netherlands,\\
           \email{\{jan,anjo\}@swi.psy.uva.nl}}

\maketitle

\addtocounter{footnote}{1}
\footnotetext{In Alexandre Tessier (Ed), proceedings of the 12th International Workshop on Logic Programming Environments (WLPE 2002), July 2002, Copenhagen, Denmark.\\Proceedings of WLPE 2002: \texttt{http://xxx.lanl.gov/html/cs/0207052} (CoRR)}

\begin{abstract}
It is next to impossible to develop real-life
applications in just pure Prolog.  With XPCE \cite{XPCE:prolog} we
realised a mechanism for integrating Prolog with an external
object-oriented system that turns this OO system into a natural
extension to Prolog.  We describe the design and how it can be applied
to other external OO systems.
\end{abstract}

\section{Introduction}

A wealth of functionality is available in object-oriented systems and
libraries.  This paper addresses the issue of how such libraries can
be made available in Prolog, in particular libraries for creating user
interfaces.  

Almost any modern Prolog system can call routines in C and be called
from C (or other imperative languages). Also, most systems provide
ready-to-use libraries to handle network communication. These
primitives are used to build bridges between Prolog and external
libraries for (graphical) user-interfacing (GUIs), connecting to
databases, embedding in (web-)servers, etc. Some, especially most GUI
systems, are object-oriented (OO). The rest of this paper concentrates
on GUIs, though the arguments apply to other systems too.

GUIs consist of a large set of entities such as windows and controls
that define a large number of operations. These operations often
involve destructive state changes and the behaviour of GUI components
normally involves handling spontaneous input in the form of
\emph{events}.  OO techniques are very well suited to handle this
complexity.

A concrete GUI is generally realised by sub-classing base classes from
the GUI system. In ---for example--- Java and C++, the language used for
the GUI and the application is the same. This is achieved by either
defining the GUI base classes in this language or by encapsulating
foreign GUI classes in classes of the target language. This situation is
ideal for application development because the user can develop and debug
both the GUI and application in one language.

For Prolog, the situation is more complicated. Diversity of Prolog
implementations and target platforms, combined with a relatively small
market share of the Prolog language make it hard to realise the
ideal situation sketched above.  In addition, Prolog is not the most
suitable language for implementing fast and space-efficient low-level
operations required in a GUI.

The main issue addressed in this paper is how Prolog programmers can
tap on the functionality provided by object-oriented libraries without
having to know the details of such libraries.  Work in this direction
started in 1985 and progresses to today.  Over these years our
understanding of making Prolog an allround programming environment has
matured from a basic interface between Prolog and object-oriented
systems (Section \ref{basic}), through introducing object-oriented
programming techniques in Prolog (Section \ref{extending}), and
finally to make it possible to handle Prolog data transparently (Section
\ref{data}). 

\section{Approaches}                            \label{sec:approaches}

We see several solutions for making OO systems available from Prolog.
One is a rigid separation of the GUI, developed in an external GUI
development environment, from the application. A narrow bridge links the
external system to Prolog. Various styles of `bridges' are used, some
based on TCP/IP communication and others on local in-process
communication. ECLiPSe \cite{eclipse:02} defines a neat generic
interface for exchanging messages between ECLiPSe and external languages
that exploits both in-process and remote communication. Amzi!{}
\cite{amzi:95} defines a C++ class derived from a Prolog-vendor defined
C++/Prolog interface class that encapsulates the Prolog application in a
set of C++ methods. The GUI part of the application is now written using
the normal guidelines of the C++ GUI library.

These approaches have advantages such as modularity and using the most
appropriate tools for the various functions of the application.
Stream-based communication is limited by communication protocol
bandwidth and latency. Whether or not using streams, the final
application consists of two programs, one in Prolog and one in an
external language between which a proper interface needs to be defined.
For each new element in the application the programmer needs to extend
the Prolog program as well as the GUI program and maintain the interface
consistency.  

For applications that require a wide interface between the application
and GUI code we would like to be able to write the application and GUI
both in Prolog. Here we see two approaches. One is to write a Prolog
layer around the (possibly extended) API of an existing GUI system
\cite{xwip,proxt}. This option is unattractive as GUI systems contain a
large number of primitives and data types. Developing and maintaining
the interface is costly. The alternative is to consider an existing OO
system where methods can be called from the C-language%
        \footnote{Or another language Prolog can communicate with. 
                  C and C++ are the de-facto interface languages.}
based on their name. Most non-C++ GUI toolkits fulfill this
requirement.%
        \footnote{The desired result can still be achieved
                  based on the C++ headers, but this is beyond
                  the scope if this article.}

In the remainder of this article we concentrate on this interesting
class of OO systems.

\section{Basic Prolog to OO System Interface}
\label{basic}

  The simplest view on an OO system is that it provides a set of
methods (functions) that can be applied to objects.  This uniformity
of representation (objects) and functionality (methods) makes it
possible to define a very small interface between Prolog and an OO
system.  All that is required to communicate is to represent
object-identity in Prolog, translate Prolog data to types (objects) of
the OO system, invoke a method and translate the result back to
Prolog.  This idea has been implemented in XPCE, and throughout this
paper we will use examples based on XPCE.

\postscriptfig{width=0.6\linewidth}{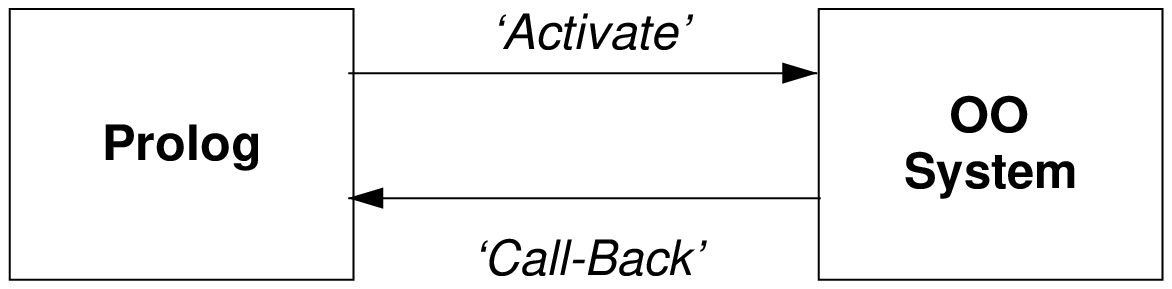}{A simple view on Prolog to
OO interface}

\subsection{Object Manipulation from Prolog}

We add four predicates to Prolog:

\begin{description}
    \predicate{new}{2}{?Reference, +Class(...Arg...)}
Create an object as an instance of \var{Class} using the given
arguments to initialise the object.  Make the resulting instance
known as \var{Reference} from Prolog.  If \var{Arg} is atomic, convert
it to a natural counterpart in XPCE (integer, float, atom).  If it is
compound, create an instance using the functor as the class name and the
arguments as initialising arguments.  Unbound variables have no sensible
counterpart in XPCE and therefore raise an exception. The example below
creates a box (graphical rectangle) with specified width and height.

\begin{verbatim}
?- new(X, box(100,100)).

X = @459337
\end{verbatim}

\noindent
    \predicate{send}{2}{+Reference, +Method(...Arg...)}
Given a reference to an object, invoke the named method with the given
arguments. The arguments are translated as with \predref{new}{2}. XPCE uses
\predref{send}{2} for methods returning either no value or a boolean
success/failure indication.  The example creates a picture (graphics
window) and displays a box at a given location.

\begin{verbatim}
?- new(P, picture),
   send(P, display(box(100,50), point(20,20))).
\end{verbatim}

\noindent
    \predicate{get}{3}{+Reference, +Method(...Arg...), -Result}
Used for methods that return a `true' result value. If the returned
value is primitive, it is converted to a Prolog integer, float or atom. In all
other cases an object reference is returned. The following code gets the
left-edge of the visible part of a window.

\begin{verbatim}
?- new(P, picture),
   get(P, visible, Visible),
   get(Visible, x, X).

P = @1072825
Visible = @957733
X = 0 
\end{verbatim}

\noindent
    \predicate{free}{1}{+Reference}
Destroy the referenced object.
\end{description}

Minimalists may claim that three of these four primitives are
redundant.  Generating an instance of a class is an operation of
the class and can thus be performed by a method of the class if classes
can be manipulated as objects (or by a `pseudo-method' of the interface
otherwise). Destroying an object is an operation on the object itself
and \predref{send}{2} is a special case of \predref{get}{3}.

We perceive the distinction between \emph{telling} an object to
perform an action without interest for the result (i.e. \emph{move} to a
location) and \emph{asking} an object to return, compute or create
something improves readability. The \predref{new}{2} and \predref{free}{1} predicates are
pleasant short-hands, also improving readability of the code.

XPCE defines access to object attributes using two methods, one for
retrieving the current value and one for setting it.  For object systems
that make a distinction between accessing attributes and methods two
more predicates are required to set and retrieving attribute values.

Life-time is an important aspect of objects and their references. Object
references (e.g.\ \objectname{1072825} in the example above) are Prolog terms acting
as an opaque handle to an object. The life-time of the object is totally
unrelated to the life-time of the Prolog reference. The user must be
aware of the object life-time rules of the object system. XPCE defines
methods to lock, unlock and free objects. In the JPL
(\url{http://sourceforge.net/projects/jpl/} project, interfacing
SWI-Prolog to Java, Paul Singleton has used Prolog atoms for
representing object-identity, synchronising object and reference
life-time using low-level hooks at both ends.

\subsection{Prolog as an Object}

  The four predicates suffice to invoke behaviour in the OO system
from Prolog.  Events and other requests from the OO system can
be placed in a queue object and Prolog can read this queue. Another
possibility is to embed Prolog in a class of the OO system and
define a method \emph{call} that takes a predicate name and a list of OO
arguments as input. These arguments are translated to Prolog the same
way as the return value of \predref{get}{3} and the method is executed by calling the
predicate. In XPCE, class \class{prolog} has a single instance with a
public reference (\objectname{prolog}). This object can, of course, also be
activated from Prolog with the following result:

\begin{verbatim}
?- send(@prolog, writeln('Hello World')).

Hello World.
\end{verbatim}

\noindent
\subsection{Portability}

The above has been defined and implemented around 1985 for the first
XPCE/Prolog system. It can be realised for any OO system that provides
runtime invocation of methods by name and the ability to query and
construct primitive data types.

\subsection{Limitations}

The basic OO interface is simple to implement and use.  Unfortunately
it also has some drawbacks as it does not take advantage of some
fundamental advantages of object-oriented programming: specialisation
through sub-classing and the ability to create abstractions in new
classes.  These drawbacks become very apparent in the context of
creating interactive graphical applications.

  Normally, application GUI-objects are created by sub-classing a
base class of the toolkit and refining methods such as \emph{OnDraw}
and \emph{OnClick} to perform the appropriate actions.  In our
interface, the implication is that we need to program the OO system.

  For XPCE, defined in the C-language, this is highly unattractive for
Prolog programmers.  For this reason we invented \emph{message
objects}.  Message objects are executed on pre-defined events, causing
some method to be called.  The following example illustrates this with
a push button calling a Prolog predicate:

\begin{verbatim}
?- new(B, button(hello,
                 message(@prolog, call,
                         writeln, 'Hello World'))).
\end{verbatim}

\noindent
This is quite satisfactory for simple cases. GUI objects however tend
to have a large number of `events' one might wish to intercept,
resulting in a large number of messages that may have to be attached
to objects.  Moreover, this type of programming occurs at the \emph{object}
rather than the \emph{class} level, thereby impairing re-use and
efficiency.

Still, even with the possibility to program in the OO system easily,
this is not always the best solution. The programmer has to work in
Prolog and the OO language at the same time. With decoupled systems
as explained before, the revenue is clean modularization. In this
case the revenues are small and the costs are large. Debugging hybrid
environments is difficult, especially if one of the languages is not
designed for this. In addition it requires programmers that can handle
both languages.

\section{Extending Object-Oriented Systems from Prolog}
\label{extending}

  If we can extend the OO system from Prolog with new classes and
methods that are executed in Prolog we have realised two major
improvements. We have access to functionality of the OO system that must
be realised by refining `virtual' methods and, with everything running
in Prolog we only need the Prolog debugger. \Figref{subclass}
illustrates our aim.

\postscriptfig{width=0.8\linewidth}{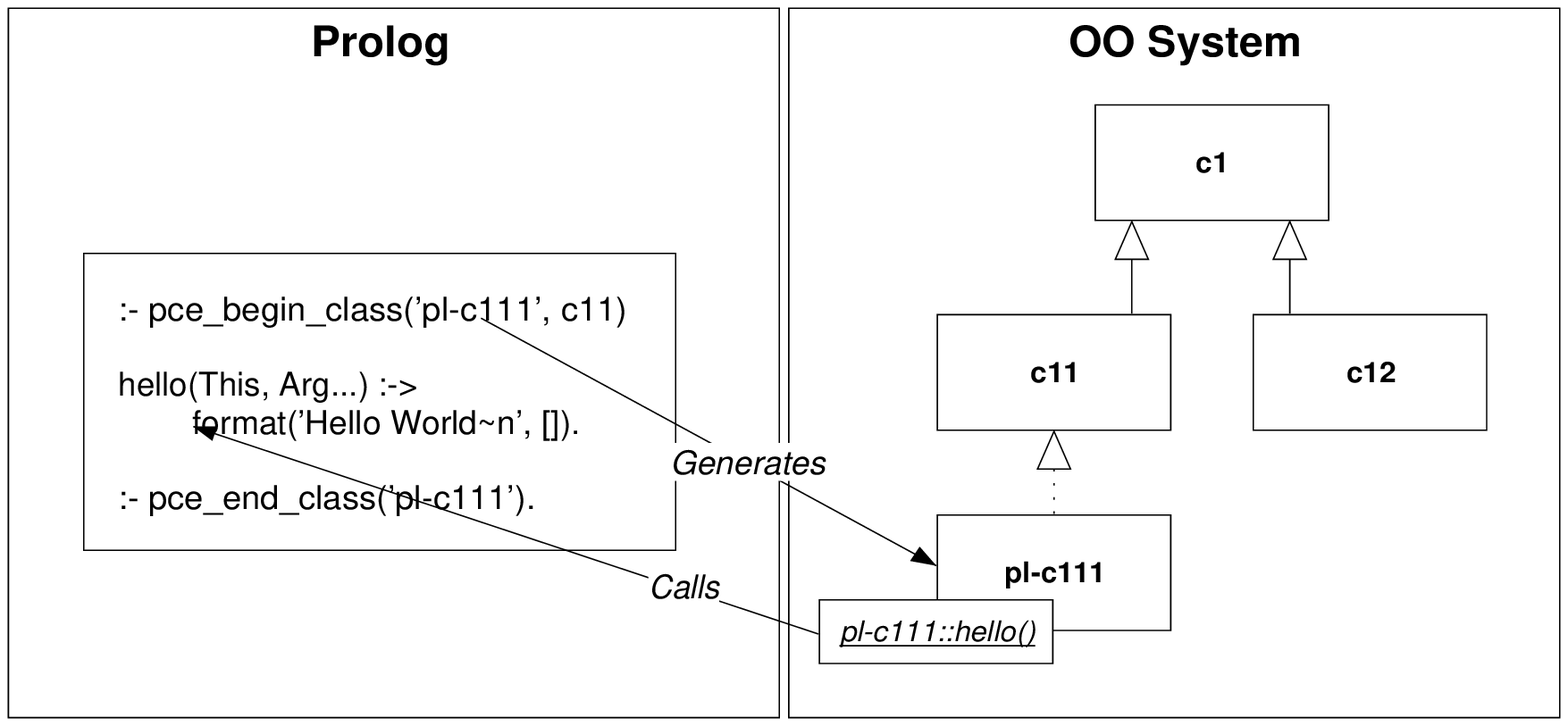}{Creating sub-classes from Prolog}

  In XPCE, methods are primary objects consisting of a name,
argument type definition and a handle to the implementation. Before
introducing sub-classing from Prolog, the implementation was a simple
pointer to the C-function realising the method. We decided to extend
this, allowing for arbitrary handles that can be interpreted by the
(Prolog) interface. On top of that we defined a specification for
writing classes in Prolog and implemented a compiler for this
specification using Prolog's \predref{term\_expansion}{2} based
macro-facility. A small example of the ``classes in Prolog'' notation:

\begin{figure}
\begin{verbatim}
:- pce_begin_class(my_box, box).

event(Box, Event:event) :->
        (   send(Event, is_a, area_enter)
        ->  send(Box, fill_pattern, colour(red))
        ;   send(Event, is_a, area_exit)
        ->  send(Box, fill_pattern, @nil)
        ;   send_super(Box, event, Event)
        ).

:- pce_end_class(my_box).
\end{verbatim}

\noindent
    \caption{Defining a class from Prolog}
    \label{fig:defclass}
\end{figure}

This program fragment defines a derived class of the base class \class{box}.
Event processing is extended such that the box is red if the pointer
(mouse) is inside and transparent (\objectname{nil}) otherwise.

XPCE is a `soft typed' language. Method arguments may have
type specifiers and if they do the system performs runtime checks on
these.  The method above requires the first argument to be an instance
of the class \class{event}.

\subsection{Implementation}             \label{sec:methodimpl}

The code from \figref{defclass} is translated into a clause for the
predicate pce\_principal: \predref{send\_implementation}{3} described below and
Prolog facts describing method and class properties. The XPCE class
and methods are created just-in-time from these facts. Representing
the class and methods in Prolog rather than creating them directly in
XPCE reduces program startup time and exploits the runtime generation
facilities of the Prolog system.

\begin{description}
    \predicate{pce\_principal:send\_implementation}{3}{+Id, +Method, +Object}
Multifile predicate created by the XPCE class compiler.  \var{Id} is a
indexable Prolog identifier of the method, \var{Method} is a term
denoting method name and arguments and \var{Object} is the receiver of
the method.
\end{description}

The method of \figref{defclass} is translated into the clause below.
Note that the head is re-arranged to allow for a fast (indexed) call,
while the body remains unchanged.

\begin{verbatim}
pce_principal:send_implementation('my_box->event', event(A), B) :-
    user:
    (   (   send(A, is_a(area_enter))
        ->  send(B, fill_pattern(colour(red)))
        ;   send(A, is_a(area_exit))
        ->  send(B, fill_pattern(@nil))
        ;   send_class(B, box, event(A))
        )
    ).
\end{verbatim}

\noindent
\subsection{Portability}                        \label{sec:portcompile}

Our implementation is based on XPCE's the ability to refine class
\class{method}, such that the implementation can be handled by a new
entity (the Prolog interface) based on a handle provided by this
interface (the atom \verb$'my_box-$\verb$>event'$ in the example).
Unfortunately very few todays object systems have this ability.

Fortunately, we can achieve the same result with virtually any OO system
by defining a small wrapper-method in the OO language that calls the
Prolog interface. Ideally, we are able to create this method on demand
through the OO systems interface. In the least attractive scenario we
generate a source-file for all required wrapper classes and methods and
compile the result using the target OO systems compiler. Even in this
unattractive setting, we can still debug and reload the method
\emph{implementation} using the normal Prolog interactive debugging
cycle, but we can only extend or modify the class and method
\emph{signatures} by running the class-compiler and restarting the
application.

\subsection{Experience}

The first version of XPCE where classes could be created from Prolog was
developed around 1992. Over the years we have improved performance and
the ability to generate compact fast starting saved states. Initially
the user community was reluctant, but the system proved valuable for us,
making more high-level functionality available to the XPCE user using
the uniform class based framework.

We improved the usability of creating classes from Prolog by making the
Prolog development environment aware of classes and methods. We united
listing, setting spy-points, locating sources and cross-referencing.
Support requests indicate that nowadays a large share of experienced
XPCE users define classes. Problems are rarely related to the
class definition system itself. User problems concentrate on how to find
and combine classes and methods of the base system that fulfill their
requirements.  XPCE shares this problem with other large and powerful
GUI libraries.

With the acceptance of XPCE/Prolog classes a new style of
application development emerged. In this style classes became the
dominant structuring factor of the application. Persistent storage and
destructive assignment make XPCE data representation attractive
alternatives to Prolog's assert/retract.

These growing expectations however exposed a problem. As long as the
XPCE/Prolog classes only extend XPCE it is natural to be restricted to
XPCE data.  As more application-oriented code is involved, so is
application-oriented data.  We needed a way to pass native Prolog data
efficiently between methods.

\section{Transparent exchange of Prolog data}
\label{data}

XPCE version 5 allows method arguments to be typed as \type{prolog}.
Whenever Prolog passes an argument to such a method, it is passed
without modification.  The passed term need not be ground and can be
(further) initialised by the called method implementation.  See
\figref{plterm}.

\postscriptfig{width=0.8\linewidth}{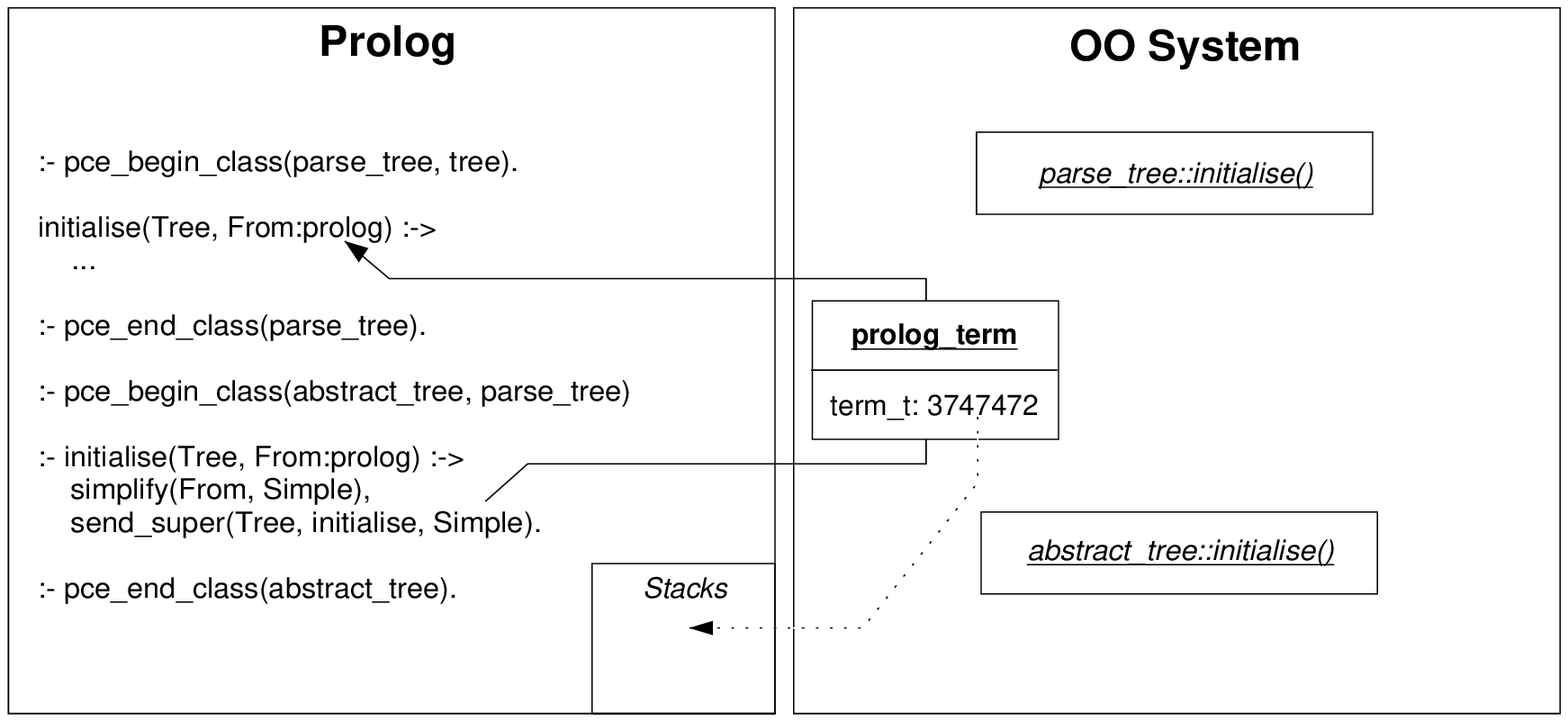}{%
        Passing Prolog terms across the OO system to other Prolog
        methods.  The term is passed `by reference'. It can contain
        unbound variables that are instantiated by the receiving
        method.}

For the technical realisation we introduced class \class{host\_data}
providing XPCE with an opaque handle to data of the \emph{host}, the
term XPCE uses for the language(s) that extend it. The Prolog interface
sub-classes this type to \class{prolog\_term}. Prolog data is opaque from
XPCE's perspective. This is sufficient, as extending XPCE itself is not
envisioned as part of normal application development.  On OO systems
where this is a likely scenario (e.g.\ Java), it is desirable to
extend the opaque handle with methods to access the Prolog term.  This
is a natural extension to the currently used implementation.

When preparing an XPCE method invocation, we check whether an argument
is of type \type{prolog}, a type that refers to \class{prolog\_term} as
well as primitive data.  If the argument is not primitive (integer,
float or atom), the interface creates an instance of \class{prolog\_term}
and stores the \ctype{term\_t} term reference in this object. Whenever an
instance of \class{prolog\_term} is passed from XPCE to Prolog, it is
presented to Prolog as a normal Prolog term.

It is natural to the XPCE programmer to be able to store data in
instance variables (slots).  XPCE uses a reference count based garbage
collector, freeing the programmer from lifetime considerations for
most objects.  The lifetime of Prolog term references however is only
guaranteed during the foreign language call where it was created. Prolog
data can only survive as a \emph{copy} in the Prolog permanent heap.

We realised a seamless integration exploiting reference count
information. After a method for which we created instances of class
\class{prolog\_term} returns, we check the reference count of the
created instances. If they are just referenced by the creator, we can
safely discard them. Otherwise, we can no longer guarantee the Prolog
term reference. The term is copied onto the permanent heap and the
reference to the term on the stacks is replaced by a reference to the
database record in the permanent heap. The Prolog record is destroyed if
XPCE's garbage collection destroys the object.  See \figref{pldata}.

\postscriptfig{width=0.8\linewidth}{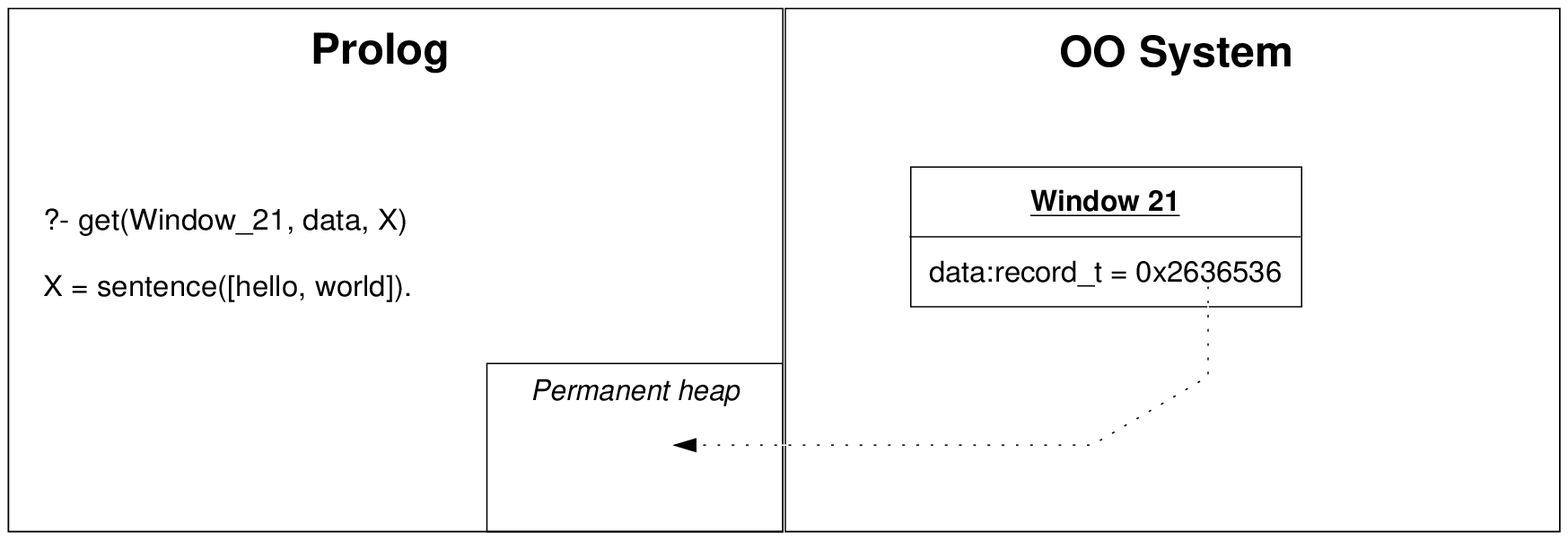}{%
        Storing Prolog data in external objects. The object contains
        a reference to a copy of the term maintained in the Prolog
        permanent heap.}

The example below creates a graphical tree where each node has a
permanent pay load of Prolog data associated. The hierarchy is created
directly from a complex Prolog data structure. After creating a
hierarchy this way, the Prolog tree as a whole is always passed \emph{by
reference} as illustrated in \figref{plterm}. Each node contains a
permanent copy of the associated pay load.

\begin{verbatim}
:- pce_begin_class(my_node, node).

variable(data, prolog, both, "Associated data").

initialise(Node, Tree:prolog) :->
        "The constructor"::
        Tree = node(Name, Data, Sons),
        send_super(Node, initialise, text(Name)),
        send(Node, data, Data),
        forall(member(Son, Sons),
               send(Node, son, my_node(Son))).

:- pce_end_class(my_node).
\end{verbatim}

\noindent
\subsection{Portability}

Modern object systems generally have a more subtle view on garbage
collection and therefore the reference-count transparent change of
status from term-reference to copy is often not feasible.  It could
be argued that such an important transition should not go by
unintentionally anyhow and it is much better to define a class that
represents a Prolog term as a reference to a copy in the Prolog
permanent heap (record).  Introduction of such a class avoids the
need for reference-counting while making the programmers intention
clear.

SWI-Prolog contains a fast C-interface to transfer terms between the
stacks and the permanent heap. This should be easily achieved in other
Prolog systems, while the slower route using \predref{assert}{1} through the Prolog
interface works in any implementation.

\subsection{Experience}

The possibility to pass Prolog data around has been introduced about 18
months ago, after which it has been used locally to become documented in
the official manual about 8 months ago.  We have little insight how much
it is used externally. Locally it is now in frequent use, clearly
simplifying and improving efficiency of methods that have to deal with
application data that is already represented in Prolog, such as the
parse tree generated by the SWI-Prolog SGML/XML parser.

\subsection{Non-deterministic methods}

Method invocation currently always passes the foreign interface, losing
non-determinism. If the implementation is in Prolog, it is desirable to
be able to exploit non-determinism. The interface may call the
method implementation directly after it detects the method is defined
in Prolog.  We have experimented with an implementation following the
following schema:

\begin{verbatim}
send(Object, Message) :-
        resolve_implementation(Object, Method, Implementation),
        Implementation.
\end{verbatim}

\noindent
For methods defined in Prolog, the implementation as shown in
\secref{methodimpl} is called immediately, retaining non-determinism
as well as last-call optimization.  For methods defined externally,
\var{Implementation} calls a foreign-language routine realising the
method.

We have not yet included this mechanism as too much existing code relies
on implicit argument type conversion provided by the interface (not
discussed here) and the fact that currently, every Prolog method
implementation implicitly ends with pruning the choice points it may
have created. Probably the best solution is to introduce additional
syntax declaring a method to be \emph{pure-Prolog}.

\section{Performance}

XPCE/Prolog message passing implies data conversion and foreign code
invocation, slowing down execution. However, XPCE provides high-level
(graphical) operations limiting the number of messages passed.
Computation inside the Prolog application runs in native Prolog and is
thus not harmed. Bottlenecks appear ---for example--- when manipulating
bitmapped images at the pixel-level or when using XPCE storage classes
as the basis for high-performance and low-level object-oriented
modeling.

Calling a typical method with two arguments defined in Prolog from
Prolog costs, depending on the nature of the arguments, between 2.4
and 5.2 $\mu{}S$ on an AMD 1600+ processor (Linux 2.4, gcc).

Table \tabref{perform} illustrates the performance on some typical
calls. The first two calls are to a C-defined built-in class involving
no significant `work' and taking no or one integer argument. The class
\class{bench} is implemented in Prolog as a subclass of the `root'
object called \class{object}. The three methods accept no arguments, an
integer argument and a Prolog term argument.

\begin{table}
\begin{center}
\begin{tabular}{|l|l|l|}
\hline
\bf Goal                & Class & Time in $\mu{}S$ \\
\hline
send(\objectname{426445}, normalise)            & area  & 1.0 \\
send(\objectname{426484}, x, 1)                 & area  & 1.5 \\
send(\objectname{426891}, noarg)                    & bench & 1.9 \\
send(\objectname{426957}, intarg, 1)                & bench & 2.3 \\
send(\objectname{427179}, termarg, hello(world))    & bench & 2.5 \\
\hline
\end{tabular}
\end{center}
    \caption{}
    \label{tab:perform}
\end{table}

\section{Threads, Events and Debugging}

XPCE/SWI-Prolog is a single-threading environment.%
        \footnote{There is a beta version of multi-threaded SWI-Prolog
                  that can be used with a thread-safe version of XPCE.
                  This version of XPCE simply serializes access from
                  multiple Prolog threads.}
Where many todays candidate object systems are multi-threaded, the
communication with Prolog often requires serialization or a Prolog
system that can cooperate with the threading model of the OO system.
This problem is not unique to the presented design, though the more
fine-grained interaction between Prolog and the object-system make
the problem more urgent.

An important advantage of the described interface is that all
application-code is executed in Prolog and can therefore be debugged and
developed using Prolog's native debugger and, with some restrictions
described in \secref{portcompile}, Prolog's incremental compilation to
update the environment while the application is running.

The Prolog debugger is faced with phenomena uncommon to the traditional
Prolog world. The event-driven nature of GUI systems causes
`spontaneous' calls.  Many user-interactions consist of a sequence of
actions each causing their own events and Prolog call-backs.  User
interaction with the debugger may be difficult or impossible during
such sequences.  For example, call-backs resulting from dragging an
object in the interface with the mouse cannot easily be debugged on
the same console.  The design also involves deeply nested control
switches between foreign code and Prolog.  The SWI-Prolog debugger
is aware of the possibilities of interleaved control and provides hooks
for presenting method-calls in a user-friendly fashion.  Break-points
in addition to the traditional spy-points make it easier to trap the
debugger at interesting points during user-interaction.  \Figref{debug}
shows the source-level debugger in action on XPCE/Prolog code.

\begin{figure}%
\centerline{psfig{file=debug.eps,width=0.8}}
\caption{SWI-Prolog source-level debugger showing break-point and
        interleaved foreign/Prolog stack context.}
\label{fig:debug}
\end{figure}

\section{Related Work}

To our best knowledge, there are no systems with a similar approach
providing GUI to Prolog.  Other approaches for accessing foreign GUI
systems have been explored in \secref{approaches}.

Started as a mechanism to provide a GUI, our approach has developed into
a generic design to integrate Prolog seamlessly with an external OO
programming language. The integrated system also functions as an object
extension to Prolog and should therefore be compared to other approaches
for representing objects in Prolog. Given the great diversity of such
systems, we consider this beyond the scope of this discussion.

\section{Conclusions}

We have presented a generic architecture for integrating OO systems with
Prolog. The design allows for extending many existing OO systems
naturally from Prolog. Using this interface the user can add new classes
to the OO system entirely from Prolog which can be used to extend the
OO system as well as for object-oriented programming in Prolog.
No knowledge of details of the OO system, such as syntax, is required.

Using dynamically typed OO systems where classes and methods can be
created at runtime through the interface (i.e.\ without generating a
source file, compiling this and loading the object code) a quick and
natural development cycle is achieved. If however, less of the OO system
is accessible at runtime, development becomes more cumbersome.

We feel that the combination of Prolog and a transparent interface to
powerful object-oriented systems suffices to create real-world
applications.

\subsection{Acknowledgements}

XPCE/SWI-Prolog is a Free Software project which, by nature, profits
heavily from user feedback and participation.  We would like to thank
Mats Carlson in particular for his contribution to designing the
representation of XPCE methods in Prolog.  We also acknowedge the
reviewers for their extensive comments and suggestions, many of which
have been used to clarify this paper.


\begin{thebibliography}{1}

\bibitem{amzi:95}
Amzi!
\newblock Integrating prolog services with c++ objects.
\newblock {\em PC AI magazine}, 9(3), May/June 1995.
\newblock {http://www.amzi.com/articles/prolog\_cpp.htm}.

\bibitem{xwip}
Ted Kim.
\newblock {\em XWIP Reference Manual, Version 0.6}.
\newblock {UCLA} Computer Science Department, 1993.
\newblock Technical Report CSD-880079.

\bibitem{eclipse:02}
Kish Shen, Joachim Schimpf, Stefano Novello, and Josh Singer.
\newblock A high-level generic interface to external programming language for
  {ECLiPSe}.
\newblock In {\em Practical Aspects of Declarative Languages}, Berlin, Germany,
  2002. Springer Verlag.
\newblock LNCS 2257.

\bibitem{proxt}
SICS.
\newblock {\em Quintus ProXT Manual}, 1998.
\newblock http://www.sics.se/isl/quintus/proxt/frame.html.

\bibitem{XPCE:prolog}
J.~Wielemaker and A.~Anjewierden.
\newblock {\em Programming in {XPCE/Prolog}}.
\newblock {SWI}, University of Amsterdam, Roetersstraat 15, 1018 WB Amsterdam,
  The Netherlands, 1992-2002.
\newblock http://www.swi.psy.uva.nl/projects/xpce/UserGuide.

\end{thebibliography}

\end{document}